\documentclass[aps,prd,twocolumn,tightenlines,floatfix,nofootinbib,amsmath,amssymb]{revtex4-1}

\usepackage{bm}
\usepackage{amsfonts}
\usepackage{latexsym}
\usepackage[latin1]{inputenc}
\usepackage{graphicx}
\usepackage{amsmath}
\usepackage{palatino}
\usepackage{mathpazo}
\usepackage[outdir=./]{epstopdf}
\usepackage{color}
\usepackage[normalem]{ulem}

\def\jnl@style{\it}
\def\aaref@jnl#1{{\jnl@style#1}}

\def\aj{\aaref@jnl{AJ}}                   
\def\apj{\aaref@jnl{ApJ}}                 
\def\apjl{\aaref@jnl{ApJ}}                
\def\apjs{\aaref@jnl{ApJS}}               
\def\apss{\aaref@jnl{Ap\&SS}}             
\def\aap{\aaref@jnl{A\&A}}                
\def\aapr{\aaref@jnl{A\&A~Rev.}}          
\def\aaps{\aaref@jnl{A\&AS}}              
\def\mnras{\aaref@jnl{MNRAS}}             
\def\prd{\aaref@jnl{Phys.~Rev.~D}}        
\def\prl{\aaref@jnl{Phys.~Rev.~Lett.}}    
\def\qjras{\aaref@jnl{QJRAS}}             
\def\skytel{\aaref@jnl{S\&T}}             
\def\ssr{\aaref@jnl{Space~Sci.~Rev.}}     
\def\zap{\aaref@jnl{ZAp}}                 
\def\nat{\aaref@jnl{Nature}}              
\def\aplett{\aaref@jnl{Astrophys.~Lett.}} 
\def\apspr{\aaref@jnl{Astrophys.~Space~Phys.~Res.}} 
\def\physrep{\aaref@jnl{Phys.~Rep.}}      
\def\physscr{\aaref@jnl{Phys.~Scr}}       
\def\commat{\aaref@jnl{Comm.~Math.~Phys.}}		
\def\science{\aaref@jnl{Science}}		
\def\cqg{\aaref@jnl{Classical Quant.~Grav.}}		
\def\jpcs{\aaref@jnl{JPCS}}					
\def\ijmpd{\aaref@jnl{Int.~J.~Mod.~Phys.~D}}			
\def\grg{\aaref@jnl{Gen.~Relat.~Gravit.}}		
\def\rpp{\aaref@jnl{Rep.~Prog.~Phys.}}		
\def\rmp{\aaref@jnl{Rev.~Mod.~Phys.}}		
\def\jpg{\aaref@jnl{J.~Phys.~G~Nucl.~Partic.}}		
\def\npa{\aaref@jnl{Nucl.~Phys.~A}}		

\begin{document}

\title{A gravitational wave afterglow in binary neutron star mergers}
\author{Daniela D. Doneva, Kostas D. Kokkotas, Pantelis Pnigouras}
\affiliation{Theoretical Astrophysics, IAAT, Eberhard-Karls University of T\"ubingen, T\"ubingen 72076, Germany}

\begin{abstract}
We study in detail the $f$-mode secular instability for rapidly rotating neutron stars, putting emphasis on supermassive models which do not have a stable nonrotating	counterpart. Such neutron stars are thought to be the generic outcome of the merger of two standard	mass neutron stars. In addition we take into account the effects of strong magnetic field and $r$-mode instability, that can drain a substantial amount of angular momentum. We find that the gravitational wave signal emitted by supramassive  neutron stars can reach above the Advance LIGO sensitivity at distance of about 20Mpc and the detectability is substantially enhanced for the Einstein Telescope. The event rate will be of the same order as the merging rates, while the analysis of the signal will carry information for the equation of state of the post-merging neutron stars and the strength of the magnetic fields.
\end{abstract}


\maketitle


\section{Introduction}
The advance of various gravitational wave detectors, such as Advanced LIGO and Virgo, KAGRA and INDIGO, makes the study of objects and processes which can emit a copious amount of gravitational radiation an extremely timely topic. One of the most promising sources in this direction  is binary neutron star mergers. In the first few tens of milliseconds after the merger the deformed and differentially rotating object can be prone to dynamical instabilities while nonradial oscillations can be excited up to a large amplitude \cite{Shibata2000,Baiotti2008,Hotokezaka2013a,Faber2012,Stergioulas2011,Takami2014}. If the mass is below a given threshold \cite{Hotokezaka2013a,Kiziltan2013,Lasky2014,DallOsso2014}, the newly formed neutron star would not promptly collapse to a black hole (BH), but instead it will enter a  quasi-stationary regime, rotating with very high frequency. At that point of the evolution dynamical instabilities are no longer present, but the rapid rotation opens a door towards secular instabilities, that can also produce copious amounts of gravitational radiation. Such an instability is the Chandrasekhar-Friedman-Schutz (CFS) instability \cite{Chandrasekhar70,Friedman78}, where the oscillation modes can be driven unstable due to the emission of gravitational radiation.

One of the most efficient emitters of gravitational waves is the $f$-mode (fundamental non-radial oscillation). In the recent years, the interest in the $f$-mode CFS instability has increased due to the  numerical and analytical  techniques \cite{Gaertig08,Gaertig10,Yoshida12,Zink10}. It was shown that, in the relativistic case, the CFS instability window (the region of the parameter space where the CFS instability overcomes the dissipative effects) can reach down to 80\% of the Kepler limit \cite{Doneva2013a} and the gravitational waves emitted by CFS unstable $f$-modes of newly born neutron stars formed after a core collapse can reach above the Advanced LIGO/Virgo sensitivity for a distance up to 10-20Mpc \cite{Passamonti2012,Passamonti2012a}. A Newtonian calculation for uniform density ellipsoids modeling newly born neutron stars after a supernova collapse can be found in \cite{Lai1995} (for a connection to the shallow decay phase in the early X-ray afterglows of gamma-ray bursts see also \cite{Corsi2009}). In this paper we present the first study of the $f$-mode CFS instability of supramassive, rapidly rotating relativistic stars, which are the outcome of binary neutron star mergers.


\section{The $f$-mode instability region} 
In order to calculate the neutron star oscillation modes we use a linear perturbation code, developed in \cite{Kruger10,Doneva2013a}, that works in the Cowling approximation, i.e. the spacetime metric is kept fixed and only the fluid variables are evolved in time. This approach simplifies the perturbation equations considerably and it gives good results for the $f$-mode frequencies and growth/damping times, not only qualitatively but also quantitatively \cite{Gaertig11,Doneva2013a,Passamonti2012}. Here we will present the methodology very briefly and a detailed presentation of can be found in \cite{Doneva2013a}. 

We adopt spherical coordinates $x^\mu=(t,r,\theta,\phi)$ and the line element of the stationary and axisymmetric spacetime induced by a rotating neutron star takes the following general form
\begin{equation}\label{eq:metric}
ds^2 = -e^{2\sigma} dt^2 + e^{2\psi}r^2 \sin^2 \theta (d\phi - \varpi dt)^2 + e^{2\mu}(dr^2 + r^2 d\theta^2)\;,
\end{equation}
where $\sigma$, $\psi$, $\varpi$ and $\mu$ are functions of $r$ and $\theta$. Due to the fact that we are working in the Cowling approximation \cite{Cowling41, McDermott83}, this is also the line element for the oscillating neutron star. The linearized perturbation equations are derived from the conservation law of the energy momentum tensor $T^{\mu\nu}$ and in the Cowling approximation we have
\begin{equation} \label{eq:Pert_EM_Tensor}
\nabla_\nu (\delta T^{\mu\nu}) = 0\;,
\end{equation}
where $\nabla_\nu$ is the covariant derivative with respect to the metric \eqref{eq:metric}.
Instead of evolving directly the perturbations for the primitive hydrodynamic quantities such as density, pressure and four velocity similar to \cite{Gaertig08}, we work directly with the components of the perturbed energy momentum tensor $\delta T^{\mu\nu}$. This approach was developed in \cite{Vavoulidis05,Vavoulidis07} and more details can be found in \cite{Kruger10,Doneva2013a}.

Calculating the oscillation frequencies can be easily done by performing a Fourier transform of the computed time series. The damping/growth  times on the other hand can not be directly extracted since we are working in the Cowling approximation where the metric perturbations and thus the gravitational-wave flux are neglected. Instead one can approximate the gravitational-wave flux using the standard multipole expansion \cite{Thorne80,Ipser91}. The total growth/damping time of the mode $\tau$ (mode amplitude $\propto e^{-t/\tau}$) will be a combination of the gravitational wave growth/damping time $\tau_{GW}$ and the damping due to dissipative processes, such as the shear and bulk viscosity, collectively denoted by $\tau_v$. Thus, we have 
\begin{equation}\label{eq:tau}
1/\tau=1/\tau_{GW}+1/\tau_v.
\end{equation} 
From the multipole expansion for the gravitational-wave flux it follows that 
\begin{equation}
1/\tau_{GW} \propto \sigma_i^{2l+1} \sigma_c,
\end{equation}
where $\sigma_i$ and $\sigma_c$ are the frequencies in the inertial and corotating frame respectively, related by the standard equation $\sigma_i = \sigma_c - m \Omega$ ($\Omega$ is the stellar rotational frequency and $m$ is the azimuthal mode number). Therefore, if $\sigma_i<0$ then $\tau_{GW}<0$ and the mode's amplitude can grow exponentially with time. In practice though the cumulative damping time $\tau$ defined by eq. \eqref{eq:tau} should be less than zero instead of $\tau_{GW}$ due to the presence of dissipative mechanisms. 

\begin{figure}
  	\centerline{\includegraphics[width=0.44\textwidth]{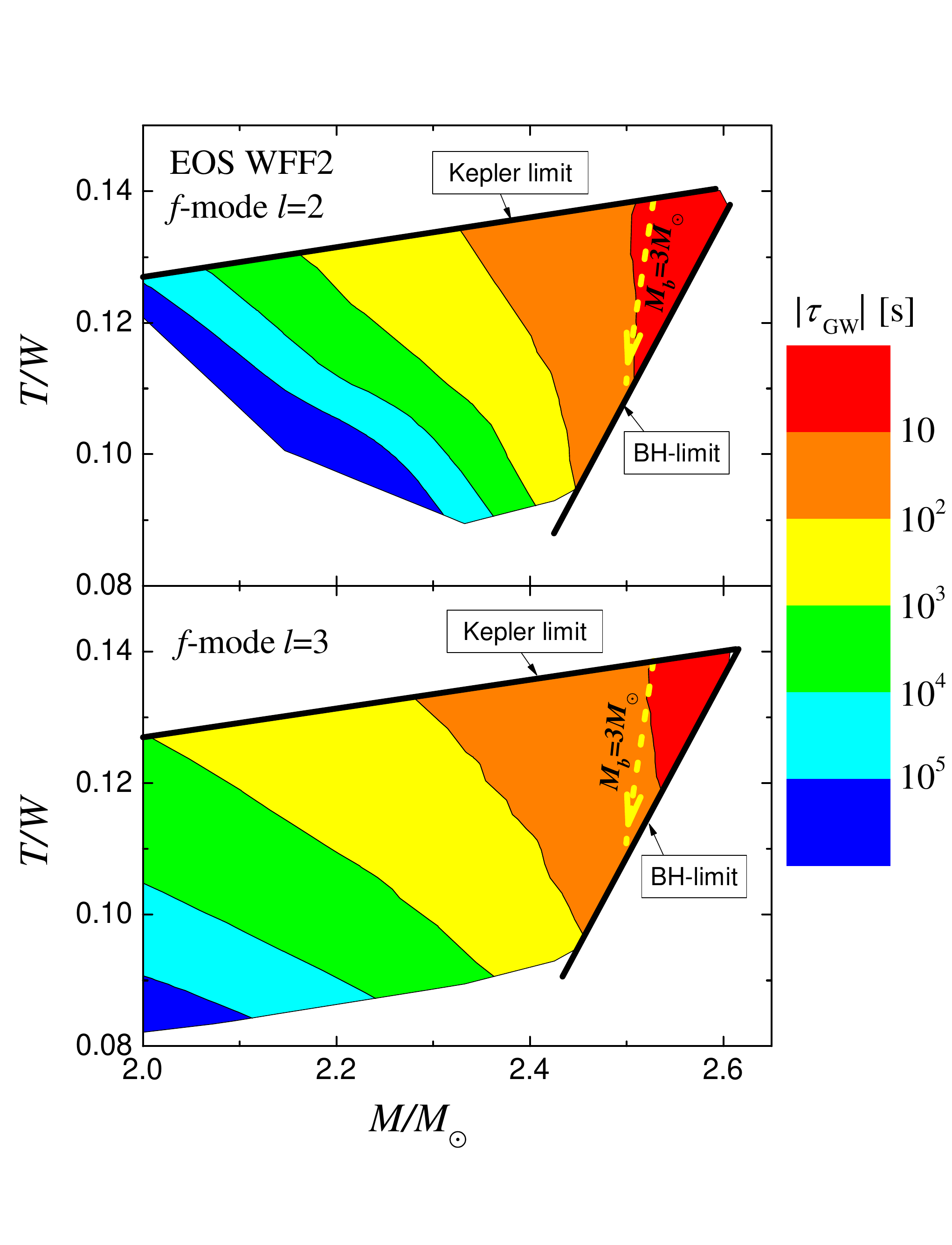}}
  	\caption{Contour plots of the growth time $|\tau_{GW}|$ for CFS unstable $f$-modes with $l=m=2$ and $l=m=3$. The chosen range of $T/W$ and the gravitational mass $M$ is in accordance with the expected merger remnant parameters. The sequence of constant baryon mass neutron stars with $M_b=3M_\odot$ is shown with a dashed yellow line.} 
  	\label{fig:fregion}
\end{figure}

Our first goal is to examine the CFS instability region and the associated growth times for supramassive stars formed after a merger. For this reason we limit our studies to models with mass $M>2M_\odot$, in contrast to the less massive models considered in \cite{Gaertig11,Doneva2013a,Passamonti2012}. In order to present the instability region for a wide range of neutron star masses and rotational rates, we present in a contour plot of the $f$-mode growth times for the CFS unstable models, i.e. the quantity $|\tau_{GW}|$ since $\tau_{GW}<0$ in this case. The results are presented in Fig. \ref{fig:fregion} for mode numbers $l=m=2$ and $l=m=3$, and the realistic equation of state (EoS) WFF2 \cite{WFF} is employed. Our rough calculations show that similar results can be obtained also for other realistic EoS, such as APR EoS \cite{Doneva2013a}. We use as independent variables on the two axes the gravitational mass of the neutron star $M$ and the ratio $T/|W|$ of the kinetic to the binding energy of the star. It is clear that, from these plots, one can not determine strictly the  models for which the CFS instability will overcome the dissipation mechanisms such as the bulk and shear viscosity, i.e. the models for with $\tau$ defined by eq. \eqref{eq:tau} is less than zero, because different models have different dissipative timescales. The bulk and the shear viscosity are strongly dependent on the temperature and our results show that for set of neutron star models we are considering, the cumulative dissipation damping time $\tau_v$ reaches up to roughly $10^7s$ for temperatures of the order of $10^9K$. For higher temperatures the CFS instability is suppressed by the bulk viscosity and for lower -- by the shear viscosity. Therefore strictly speaking if $|\tau_{GW}|<10^7s$ we will have a nonvanishing instability window. In practice though  $|\tau_{GW}|$ should be at least one  or two orders of magnitude smaller in order to have an appreciable instability region (see also \cite{Doneva2013a,Gaertig11,Passamonti2012}).

Several important conclusions can be drawn from this figure. First of all, one can see that growth times lower than $100{\rm s}$ exist for a large region of the parameter space. Moreover, the typical neutron stars formed after the merger will rotate very fast with masses about $2.5M_\odot$ \cite{Hotokezaka2013a,Kiziltan2013,Lasky2014,DallOsso2014}. Therefore they occupy the region with the shortest growth times in Fig.  \ref{fig:fregion} (of the order of tens of seconds). This is orders of magnitude smaller than the typical growth times for the less massive relativistic neutron stars formed after a core collapse \cite{Passamonti2012} and would naturally lead to a very rapid development of the CFS instability. Such high mass neutron stars often do not have a stable nonrotating limit, but instead they will collapse to a BH after losing a significant portion of their angular momentum. The EoS we are considering allows for stable rotating equilibrium solutions with such high masses, but one should keep in mind that this might not be true for some softer EoS that lead to a prompt collapse to a BH after the merger.

$f$-modes with lower values of $l$ reach shorter growth times, but they are CFS-unstable for a smaller region of the parameter space. Hence, for lower mass neutron stars, the $l=m=3$ and $l=m=4$ modes have much larger instability windows compared to the $l=m=2$ case and they dominate the gravitational wave emission \cite{Passamonti2012}. But the $l=m=2$ modes can become easily unstable for larger masses, with shorter growth times compared to the higher-$l$ modes as Fig. \ref{fig:fregion} shows. Therefore, the $l=m=2$ modes will give the dominant contribution for the supramassive neutron stars formed after a merger. In the following sections we will concentrate exactly on this case. 

\section{Instability evolution and gravitational wave emission}
In this section we will study the characteristic features of the evolution through the instability region for supramassive neutron stars formed after a binary merger and determine how strong the emitted gravitational wave signal is. We perform the time evolution following the methodology described in \cite{Passamonti2012}. We are not going to present in detail the evolution equations since they are quite lengthy and there are not significant deviation from \cite{Passamonti2012}. Instead we will discuss here only the main points and assumptions.

 The independent quantities that are evolved are the amplitude of the mode, the angular momentum and the temperature. The procedure can be described as follows. The $f$-mode amplitude initially grows exponentially due to the CFS instability. After reaching a given saturation amplitude nonlinear mechanisms begin to drain energy from the mode and the amplitude remains practically constant until the star exits the instability window or collapses to a BH. The angular momentum loss on the other hand is due to two competing mechanisms -- the gravitational wave emission of the oscillating star and the magnetic dipole radiation. The temperature evolution is governed by neutrino cooling and reheating due to  shear viscosity. Here, we neglect the initial hot phase of the neutron star life when the temperature effects are non-negligible, for two main reasons. First, for temperatures a few times $10^{10}{\rm K}$ the bulk viscosity can be very strong, thus completely suppressing the CFS instability. Second, the cooling timescale is of the order of tens of seconds, which is much shorter than the timescale of the evolution though the instability region (minutes or hours). In our calculations we also do not take into account differential rotation, because it will cease to exist in the first few seconds or even milliseconds after the merger (see for example \cite{Hotokezaka2013a,Shapiro2000}) and, therefore, it will have a negligible effect on the CFS instability evolution.

Concerning the initial angular momentum of the models we should keep in mind the following. There are processes in the early life of the post-merger neutron stars that can emit substantial amount of angular momentum during the differentially rotating phase \cite{Shibata2000,Baiotti2008,Hotokezaka2013a,Faber2012,Stergioulas2011,Takami2014}. But on the other hand a newborn differentially rotating star can sustain much larger angular momentum compared to its uniformly rotating counterpart. The precise amount of angular momentum emitted during this phase is still unknown and that is why we assume, as an estimate, that the secular quasi-stationary evolution of the uniformly rotating neutron star starts from a near-Kepler frequency. For the models we consider the star collapses to a BH after losing about 10-20\% of its angular momentum $J$, that corresponds to $\Delta J = 4-8\times 10^{48}{\rm g\,cm^2/s}$.

Choosing the saturation amplitude is also an important point that deserves a detailed analysis. The only studies about the $f$-mode saturation amplitude up to now are \cite{Kastaun10} and \cite{PK2014}. In \cite{Kastaun10}, the nonlinear evolution of perturbations with an arbitrarily large amplitude was studied and the threshold for the onset of nonlinear damping was determined for a variety of models. For this purpose, a general relativistic nonlinear hydrodynamics code in conjunction with a fixed spacetime (Cowling approximation) was used and different damping mechanisms, such as wave breaking and shock formation, were investigated. The results in \cite{Kastaun10} represent an upper limit on the saturation amplitude coming from damping mechanisms operating on dynamical timescales, since the simulation time is limited to a few tens of milliseconds. Another efficient channel for mode energy transfer is the nonlinear mode coupling. This normally involves secular timescales that can not be studied with nonlinear hydrodynamic codes. Instead, a different code was developed in \cite{PK2014}, in the Newtonian approximation, with the main objective to study the possible nonlinear three-mode couplings (coupling between an $f$-mode and two other modes) and the corresponding transfer of energy, in a manner similar to the $r$-mode case \cite{Schenk02,Arras2003,Brink2005,Brink2004,Brink2004a,Bondarescu2009}. At the end, the actual saturation amplitude of a mode will be the minimum of the values obtained in \cite{Kastaun10} and \cite{PK2014}.

The results in \cite{Kastaun10} showed that for all of the studied models and types of modes, both axisymmetric and nonaxisymmetric, the saturation amplitude is above $10^{-6}Mc^2$. For the $l=m=2$ $f$-modes, which we are interested in the present paper, the saturation amplitude reaches $10^{-4} - 10^{-5} Mc^2$. The results in \cite{PK2014} on the other hand are strongly dependent on the rotational frequency and the temperature of the star since the possible couplings of the $f$-mode, that lead to efficient energy transfer, can change significantly with the change of the neutron star parameters. The upper limit of the $l=m=2$ $f$-mode saturation amplitude for supramassive models, similar to the ones presented in the current paper\footnote{In \cite{PK2014}, Newtonian approximation is used due to the high complexity of the problem. Thus, one can not obtain the same supramassive models as the ones used in the present paper, where the relativistic effects are well pronounced. Instead, some approximations, discussed in detail in \cite{PK2014}, are introduced. Of course, this can also lead to a change of the upper limit of the saturation amplitude, compared to the general relativistic case.}, is of the order of $10^{-6} Mc^2$. Therefore, as an estimate, consistent both with \cite{Kastaun10} and \cite{PK2014}, we will work with a saturation amplitude $\alpha_{\rm sat}^f=10^{-6} Mc^2$. The dependence of our results on $\alpha_{\rm sat}^f$ will be presented in detail further on.

In what follows we will study in detail the produced signal-to-noise ratio (S/N)  by the CFS unstable $f$-modes for three models. The effect of strong magnetic field and the presence of unstable $r$-modes will be also taken into account, since they can drain a substantial amount of the stellar angular momentum that will eventually reduce the observed $f$-mode S/N ratio.

Let us quickly outline the methodology for calculating the S/N ration (see for example \cite{Owen1998}). It can be easily obtained once we know the gravitational wave strain produced by the neutron star oscillations $h(t)$ and the power spectral density of the noise of the detector $S_h(\nu)$. One can calculate $h(t)$ using the mass and the current multipole moments \cite{Thorne80} \footnote{For the $f$-modes the dominant contribution comes from the mass multipole moment.}. The Fourier transform of $h(t)$ is given in the stationary phase approximation by
\begin{equation}
|h(t)|^2 = |{{h}} (\nu)|^2 \left|\frac{d\nu}{dt}\right|,
\end{equation}
where $\nu$ is the frequency. 

We will present the optimal value for the S/N ration that can be obtained by matched filtering. In this case   we have
\begin{equation}
\left(\frac{S}{N}\right)^2 = 2 \int_0^\infty \frac{|{h}(\nu)|^2}{S_h(\nu)} d\nu
\end{equation}
where ${h}(\nu)$ is averaged over the orientation of the source and its location on the detector's sky.

\begin{figure}
	\centerline{\includegraphics[width=0.5\textwidth]{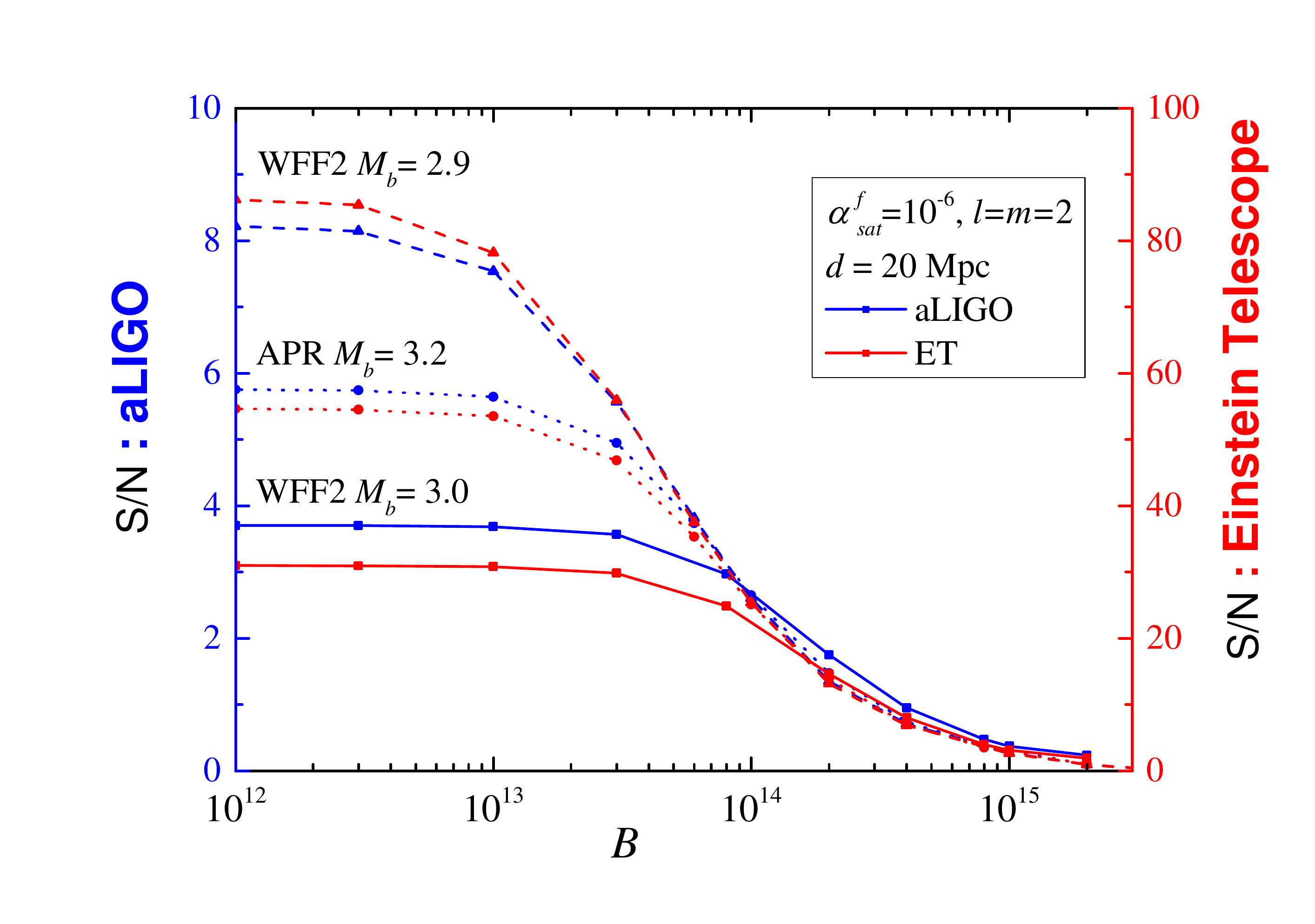}}
	\caption{The signal-to-noise ration for Advanced LIGO (blue lines, left y-scale) and ET (red lines, right y-scale) as a function of the dipole component of the magnetic field on the surface $B$. We consider the $l=m=2$ $f$-modes of three neutron star models: a) EOS WFF2 with baryon mass $M_b=2.9M_\odot$ and gravitational mass at the Kepler limit $M_K=2.45 M_\odot$, b) EOS WFF2 with $M_b=3.0M_\odot$ and $M_K=2.53M_\odot$, c) EOS APR  with $M_b=3.2M_\odot$ and $M_K=2.70M_\odot$. The saturation amplitude is $\alpha_{sat}^{f}=10^{-6}$ and the distance to the source is $d=20{\rm Mpc}$. } 
	\label{fig:hlm}
\end{figure}

In Fig. \ref{fig:hlm} we plot the S/N ratio \cite{Jaranowski2009,Owen1998,Owen2002} as a function of the dipole component of the magnetic field on the surface, produced by the $l=m=2$ $f$-mode for three neutron star models with two different equations of state, baryon masses around  $3.0 M_\odot$ and radii in the range of $10-13\;{\rm km}$. These models do not have a stable nonrotating limit and they are typical representatives for a merger outcome. It is assumed that the distance to the sources is $d=20 {\rm Mpc}$. The S/N both in the case of the Advanced LIGO (blue lines, left $y$-axis) and Einstein Telescope (ET) (red lines, right $y$-axis) are shown. The frequency range of the signal is roughly a) 930--1030 Hz, b) 360--810 Hz and c) 630--780 Hz for the three models respectively.

The most important observation is that, for a dipole component of the magnetic field on the surface $B$ up to roughly $10^{14} {\rm G}$,  the S/N ratio of the gravitational wave signal reaches above eight for Advance LIGO at distance of 20Mpc, and it is significantly enhanced for the ET (one order of magnitude). For larger values of the surface magnetic field $B$, the magnetic dipole radiation drains rotational energy from the star very fast. Therefore less energy is emitted via gravitational waves and the S/N ratio reduces significantly. The question is how strong magnetic fields can be produced during the neutron star merger. Shortly after the merger, the supramassive neutron star will be hot and differentially rotating, and different magnetohydrodynamical instabilities, such as the Kelvin-Helmholtz and the magneto-rotational instability, will be operating \cite{Siegel2013,Kiuchi2014,Rembiasz2015}. Due to limitations of the simulations the problem is not fully studied yet, but it is expected that these instabilities can amplify the initial magnetic field by 2-4 orders of magnitude. Therefore, if the initial magnetic field of the merging neutron stars is of the order of $10^8-10^{12} {\rm G}$, which is justified by the observations, the final value of the dipole component of the magnetic field on the neutron star surface can be of the order of $10^{13}-10^{14} {\rm G}$ \cite{Siegel2013,Rembiasz2015}, although the average value of the combined toroidal and poloidal field can be considerably larger. 

\begin{figure}
	\centerline{\includegraphics[width=0.5\textwidth]{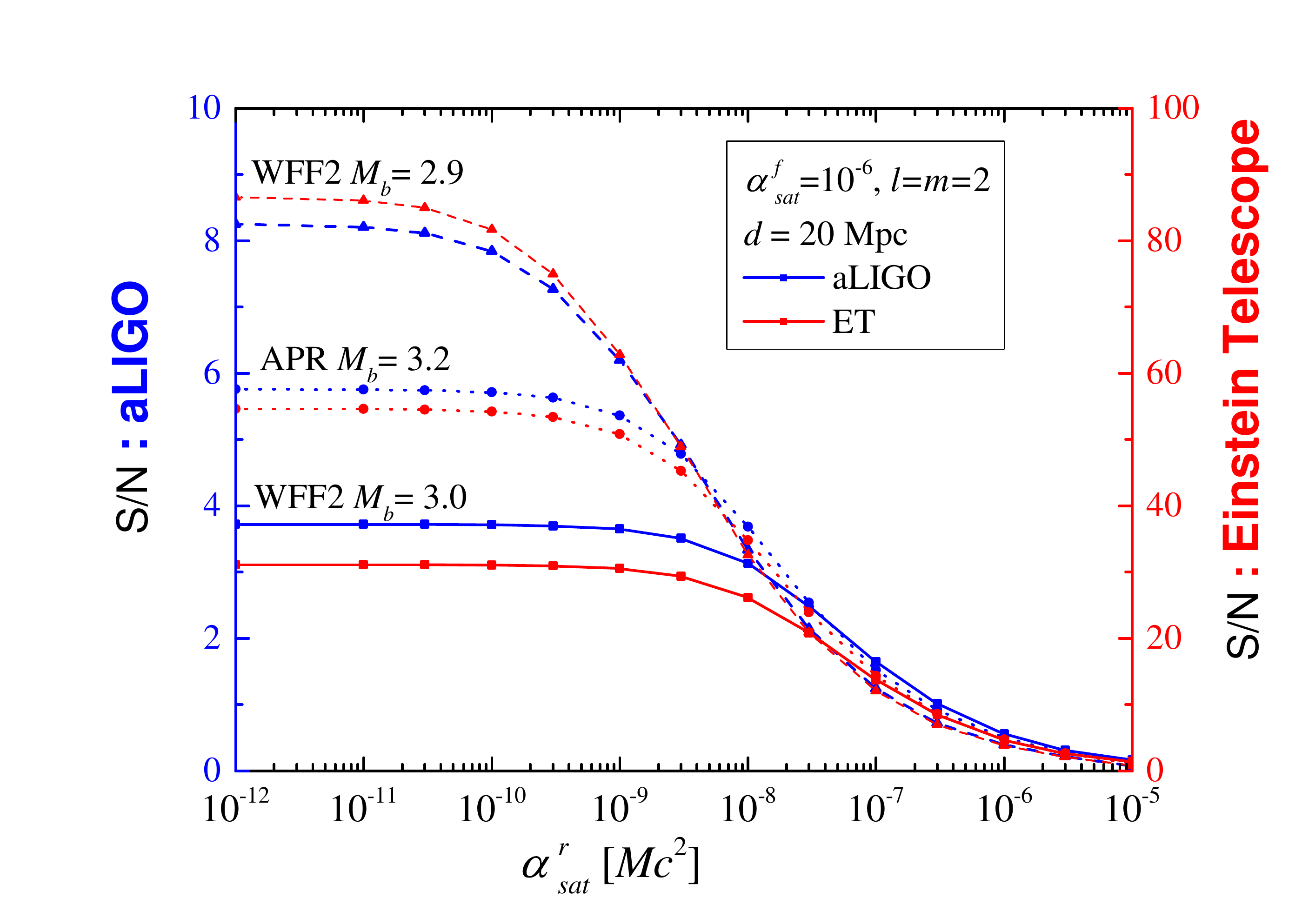}}
	\caption{The signal-to-noise ration for Advanced LIGO and ET as a function of the $r$-mode saturation amplitude. The notations are the same as in Fig. \ref{fig:hlm}.} 
	\label{fig:rmode}
\end{figure}

As a final ingredient in our study we also took into account the presence of unstable $r$-modes by considering two-mode evolution, following the methodology in \cite{Passamonti2012}. This would affect the $f$-mode evolution in a manner similar to the presence of a magnetic field -- the unstable $r$-modes can drain a substantial amount of rotational energy from the star depending on their saturation amplitude $\alpha_{sat}^r$, which shortens the evolution time. In Fig. \ref{fig:rmode} we plot the $f$-mode S/N ratio as a function of $\alpha_{sat}^r$. As one can see for $\alpha_{sat}^r$ below roughly $10^{-8}$ the S/N ration still has appreciable values both for Advance LIGO and the ET. Of course this upper limit of $\alpha_{sat}^r$ is obtained for $\alpha_{sat}^f=10^{-6}$, but a general observation is that if the $r$-mode saturation amplitude is roughly two orders of magnitude smaller than $\alpha_{sat}^f$, the unstable $r$-mode would not affect significantly the $f$-mode evolution and the corresponding S/N ration. The studies for the nonlinear mode coupling of $r$-modes \cite{Schenk02,Arras2003,Brink2005,Brink2004,Brink2004a,Bondarescu2009} suggest that $\alpha_{sat}^r \sim 10^{-8}-10^{-10}$ (in units $M c^2$). If strong magnetic field is present, the influence of the $r$-modes can be even smaller \cite{Rezzolla2000}. 

A natural question to ask is how the results would be affected by a change in the saturation amplitude. Since the magnetic dipole radiation is proportional to the square of the magnetic field $B$ \cite{Passamonti2012}, the results shown in Fig. \ref{fig:hlm} will remain nearly unaltered if we change $\alpha_{sat}^f$ and rescale the magnetic field as $\sqrt{(\alpha_{sat}^f/10^{-6})}\,\,\,B$. On the other hand the loss of angular momentum is proportional to the saturation amplitude and the results in Fig. \ref{fig:rmode} will be quite similar if we change $\alpha_{sat}^f$ and rescale the $r$-mode saturation amplitude as $(\alpha_{sat}^f/10^{-6})\,\,\alpha_{sat}^r$.

Let us discuss the potential detectability of the emitted gravitational wave signal. As one can see from Figs. \ref{fig:hlm} and \ref{fig:rmode}, the S/N ratio reaches above eight for the Advance LIGO and it is enhanced by approximately one order of magnitude for the ET. These results are obtained in the case of matched filtering that gives only an upper limit for the detection. In practice one should have S/N ratio above at least 5 in order to detect the signal \cite{Cutler1993,LIGO2013}. We will adopt this value as a detection threshold similar to other studies \cite{Corsi2009}. Therefore the signal can be detected by Advanced LIGO from sources as far as 20Mpc that is roughly the distance to the Virgo cluster. The event rate is comparable with the one for binary neutron star mergers. Unfortunately it is quite low at 20Mpc and we will probably not be able to see an event in the Advance LIGO era. Taking into account that the characteristic strain and the S/N ratio are inversely proportional ot the distance to the source, we will be able to detect events with the Einstein Telescope at distance even above 200Mpc and it is expected that we will have at least a few events per years. 

We should note that the detection can be facilitated by the following facts. Most probably a gravitational wave signal emitted from the same source during the inspiral phase or shortly after the merger will be also detected.  Some neutron star parameters, like the mass, can be determined independently from  these observations. In this case one can give a good estimate for the time when the signal from CFS unstable $f$-modes should appear and the extracted  parameters can be used to choose a narrower subset of theoretically modeled templates for the $f$-modes instability.


\section{Concluding discussion} 
In the present paper we have addressed the possibility of a detectable afterglow of  strong gravitational radiation emitted shortly after a neutron star merger. The mechanism is based on the development of the $f$-mode secular instability in the newly formed, supramassive and extremely fast rotating neutron stars. Our studies show that for these objects the instability has the shortest growth times in the whole parameter range of neutron star masses and rotational rates, which makes them important emitters of gravitational radiation. In order to prove this hypothesis, we followed the evolution of a single neutron star from its birth after the merger, up until it loses a significant portion of the angular momentum and collapses to a BH. Our calculations show that the gravitational wave signal of the unstable $f$-mode will be potentially detectable from a source located at 20Mpc for moderate or even strong magnetic fields by Advanced LIGO. The sensitivity of the Einstein Telescope in the relevant frequency range is about one order of magnitude larger than Advanced LIGO so we can detect events at about 200Mpc. As far as the $r$-modes are concerned, if the $r$-mode saturation amplitude is roughly two orders of magnitude lower than the $f$-mode saturation amplitude, they will have small effect on the $f$-mode evolution. 

The event rate will be comparable with the event rate of binary neutron star mergers. Since it is quite low at 20Mpc, the gravitational waves emitted by post-merger supramassive neutron stars, due to the secular CFS instability can hardly be detectable by Advanced LIGO but it can give at least a few event per year for the ET.   The detection of such waves will provide unique information for the post-merger EoS of the supramassive neutron stars. The reason is that only a certain number from the known EoSs allows for the creation of these objects, while the frequency, the evolution time and the amplitude of the emitted gravitational waves will critically depend on the stiffness of the final EoS. 



DD acknowledges support from the European Social Fund and the Ministry Of Science, Research and the Arts Baden-W\"urttemberg, and the Bulgarian NSF Grant DFNI T02/6. DD is grateful to S. Yazadjiev for helphul discussions and suggestions. We also thank J. Friedman, K. Glampedakis, L. Rezzolla, D.I. Jones and N. Stergioulas for careful reading of the manuscript and helpful suggestions.

\bibliographystyle{apsrev4-1}
\bibliography{references}

\end{document}